\documentclass[draft]{agujournal2019}
\usepackage{url} 
\usepackage{lineno}
\usepackage{soul}
\usepackage{siunitx}
\usepackage{amsmath}
\usepackage{booktabs}

\draftfalse

\journalname{JGR: Machine Learning and Computation}

\begin{document}

\title{HEIMDALL: a grapH-based sEIsMic Detector And Locator for microseismicity}

\authors{Matteo Bagagli$^1$}
\authors{Francesco Grigoli$^1$}
\authors{Davide Bacciu$^2$}

\affiliation{1}{Earth Sciences Department, University of Pisa, Via Santa Maria, 53, 56126 Pisa}
\affiliation{2}{Computer Science Department, University of Pisa, Largo Bruno Pontecorvo, 3, 56127 Pisa}

\correspondingauthor{Matteo Bagagli}{matteo.bagagli@dst.unipi.it}

\begin{keypoints}
\item Spatio-temporal graph-based neural network for microseismicity monitoring, both for offline and near-real-time applications.
\item Full pipeline integration for seismic phase picking, association, and final event location.
\item Application to Hengill geothermal site (Iceland) with improved detection capability and low rates of false and missed events.
\end{keypoints}

%
%

\begin{abstract}
In this work, we present a new deep-learning model for microseismicity monitoring that utilizes continuous spatiotemporal relationships between seismic station recordings, forming an end-to-end pipeline for seismic catalog creation. It employs graph theory and state-of-the-art graph neural network architectures to perform phase picking, association, and event location simultaneously over rolling windows, making it suitable for both playback and near-real-time monitoring. As part of the global strategy to reduce carbon emissions within the broader context of a green-energy transition, there has been growing interest in exploiting enhanced geothermal systems.  Tested in the complex geothermal area of Iceland’s Hengill region using open-access data from a temporary experiment, our model was trained and validated using both manually revised and automatic seismic catalogs. Results showed a significant increase in event detection compared to previously published automatic systems and reference catalogs, including a $4M_w$ seismic sequence in December 2018 and a single-day sequence in February 2019. Our method reduces false events, minimizes manual oversight, and decreases the need for extensive tuning of pipelines or transfer learning of deep-learning models. Overall, it validates a robust monitoring tool for geothermal seismic regions, complementing existing systems and enhancing operational risk mitigation during geothermal energy exploitation.
\end{abstract}

\section*{Plain Language Summary}
Geothermal energy -heat drawn from deep within the Earth- can provide constant, low-carbon electricity, making it an attractive option in the shift toward cleaner energy sources. Yet pumping fluids underground to unlock this heat can sometimes trigger old faults and cause small earthquakes. To keep people and infrastructure safe, operators must track these "micro-earthquakes" in real-time. We built a new deep-learning model that does exactly that. Instead of performing sequential steps, our algorithm simultaneously monitors the ground-motion recordings from multiple sensors and, in a single sweep, determines (1) whether and when an earthquake begins, (2) which stations detected the same event, and (3) the location of the quake beneath the surface. We tested the method in Iceland’s Hengill geothermal area, a geologically complex region with both natural and human-induced earthquakes. Using freely available data, the system spotted many more genuine events and far fewer false ones than existing automatic tools during two busy periods in 2018-2019. Because it requires little manual fine-tuning, the approach can help operators and scientists maintain better oversight of geothermal projects, fill gaps in earthquake catalogs, and make faster decisions to reduce seismic risk while advancing clean-energy exploitation goals.

%
%

\section{Introduction}
The analysis and characterization of microseismicity, particularly in relation to induced seismicity, present significant challenges for seismologists. Induced seismicity, often resulting from anthropogenic activities such as hydraulic fracturing and geothermal energy extraction, can manifest as low-magnitude seismic events that are difficult to detect and characterize using traditional seismic monitoring techniques \cite{grigoli2017}. The complexity of these events, coupled with the considerable background noise typically found in continuous data, complicates the detection of genuine seismic signals. Furthermore, the rapidly growing volume of seismic data collected by local and regional seismic networks worldwide, including temporary networks, raises questions about the effectiveness of older detection algorithms compared to newer techniques. Indeed, enhancing computational performance becomes even more crucial in near-real-time or network monitoring scenarios (i.e., early warning systems, EWS)
Recent advancements in deep learning and machine learning methodologies have shown promise in addressing these challenges, and the development of these methodologies has rapidly progressed in observational seismology. Many picking algorithms and seismic event detection techniques have been developed using machine learning approaches \cite{enescu1996, dai1995,wang1995, mousset1996, gentili2006, beyreuther2012, kong2018}. Most of these algorithms leverage the power of deep learning networks (LeCun et al. 2015), using raw seismograms or minimal pre-processing stages \cite{liao2021, mousavi2019, mousavi2020, mousavi2022, muenchmeyer2022, ross2018, ross2018a, soto2021, woollam2019, yu2022, zhu2019}.
Many of the algorithms mentioned above are developed using some Convolutional Neural Network layers \cite{lecun1995} as relatively simple Encoder-Decoder sequences \cite<e.g.,>{ross2018}. Others use more sophisticated U-net representations \cite<i.e.,>{zhu2019}, or a combination of different layers, such as convolutional and recurrent units \cite<i.e.,>{mousavi2019}, or even include additional attention mechanisms \cite{mousavi2020}.
All of these approaches, however, are station-based algorithms that lack generalization across the entire seismic network. Therefore, additional methods for event detection and phase picking have been explored, incorporating some aggregation modules \cite{vandenende2020, xiao2021, feng2022} that more effectively identify weak seismic signals amidst background noise by aggregating feature embeddings from station pairs using a cross-correlation module \cite{xiao2021}. Due to limited communication between station pairs, the consistency of waveforms across multiple stations is not fully examined.
Another possible approach is to leverage the attention mechanisms that are widely implemented in modern architectures (i.e., transformers) and used in many large language models (LLMs). When applied to time series analysis, they typically perform better at extracting relevant temporal features within the analyzed time window. Currently, the most commonly used network that implements a soft-attention mechanism at a deeper layer level is EQT \cite{mousavi2020}, which is applied station-wise for standard event detection, phase picking, and event labeling. One solution to establish communications between multiple stations is to utilize an advanced graph neural network architecture to leverage a more realistic message passing through the Graph Convolution Network (GCN).
For example, Edge-PHASE \cite{feng2022}, which utilizes Graph Ordinary Differential Equations (ODEs), highlights the inherent complexities in flow data, which can exhibit both long-range spatial dependencies and intricate temporal patterns, albeit only for phase-picking.
One of the latest deep-learning paradigm are represented by the Graph Neural Network \cite<GNN,>{sperduti1997}. This approach works with graphs, which are an ensemble of nodes (stations) and edges (connections) that define their geometry and relations. This graph approach has been applied in several fields, including molecular drug design \cite{marchant2020, jumper2024}, finance \cite{liu2021}, and urban infrastructure \cite{cui2019}.
The primary difference between convolutional neural networks (CNNs) and graph neural networks (GNNs) lies in their distinct data structures and information-processing methods. CNNs operate on regular grids, most commonly images, relying on uniform spatial relationships among pixels. Their filters slide over structured inputs to capture local features, utilizing techniques such as pooling to reduce dimensionality while retaining key information and efficiently learning hierarchical representations across multiple layers. GNNs, in contrast, extend convolution to graph-based data, where nodes and edges represent entities and their connections. Each node aggregates information from its neighbors through message passing, capturing complex relational patterns, and accommodating various edge types. This flexibility makes them suitable for applications ranging from social networks to molecular modeling. A defining advantage of GNNs is their ability to handle irregular or non-Euclidean structures and integrate edge features, providing richer insights than grid-based approaches.

Yet GNNs also face challenges. Over-smoothing can arise when expanding layers cause node representations to converge, diminishing distinctiveness. Additionally, large graphs demand substantial computational resources due to repeated message passing. Ultimately, each architecture excels in specific contexts: CNNs are well-suited for structured, grid-like data, leveraging localized convolutions and spatial hierarchies, while GNNs effectively address the intricacies of graph-structured data with dynamic connectivity. The ideal choice depends on the data characteristics and the task requirements. In observational seismology, the usage of GNN for phase association or relocation \cite{mcbrearty2023, mcbrearty2024} indeed leverages the network geometry and inter-station relations. Previous similar application for seismic source characterization was made by \citeA{vandenende2020}.
All of these algorithms and codes, however, were developed for single-use case scenarios, including picking, phase association, or event location and detection. Recent frameworks like QSeek \cite{isken2025} and QuakeFlow \cite{zhu2022} merged the complete pipelines of using ML-based seismic pickers, phase associates, and locators in a sequential manner, where each step of the algorithms undergoes fine-tuning of hyperparameters and often relies on pre-trained models.

It is important to stress, though, that the picking-association-location problem is intrinsically linked, and there is substantial interdependence among each part in play \cite<i.e.,>{kissling1988}. An initial step towards embracing this integrated philosophy was proposed by \citeA{si2024}, introducing a unified framework for regional monitoring applied to the Japan and Ridgecrest datasets. However, their architecture relies on an indirect picking mechanism conditioned on depth and offset estimates inferred through a learned physical propagation model, ultimately requiring triangulation for final localization. This not only introduces architectural complexity—requiring intermediate representations like waveform alignment and time shifts—but also limits flexibility in processing continuous data, as it assumes an event is present in every window.
Therefore, in this work, we introduce HEIMDALL, a graph-based seismic detector and locator, aiming to exploit the spatiotemporal relation that naturally describes wave propagation and the recording sequence at neighboring stations of a seismic network, for simultaneous phase-picking, association, and event location. Our approach is explicitly data-driven, fully end-to-end trainable architecture, with lightweight feature extractors, node-based processing, and direct prediction of interpretable 2D likelihood maps that allow robust event identification and localization without relying on fixed geometrical post-processing. We designed this software to build a tailored encoder-decoder architecture for microseismic applications (see the Method section).

\begin{figure}
\centering
  \includegraphics[width=0.7\columnwidth]{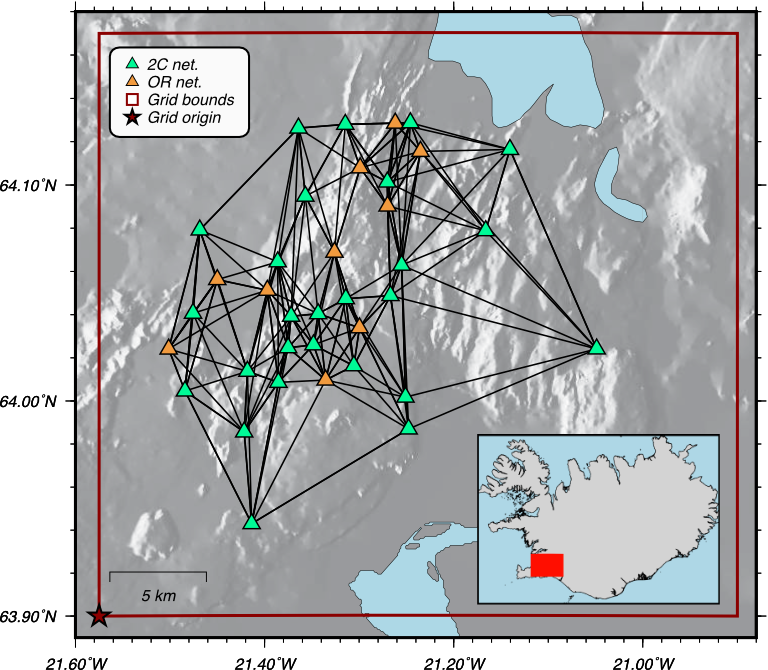}
  \caption{Map of the study area. Triangles represent the recording seismic stations used in our work, the connections represent the edges of our graph. Red square delimits the location grid area. The red star marks the grid origin point for the planar-geographical coordinates translation.}
  \label{mapgraph}
\end{figure}

We apply this algorithm to the Hengill region in southwest Iceland, approximately \SI{30}{\km} East of Reykjavík, using the COSEISMIQ project’s temporary experiment dataset \cite{grigoli2022}. After training our model with a subset of representative seismic events in the area, we tested our method on a single-day sequence (2019-02-03) and a full month (December 2018) of seismic data. We then compare our catalogs with the manually compiled one by the responsible local agency (ISOR), and automatically compiled catalogs available from the COSEISMIQ project itself \cite{grigoli2022} and one obtained from the MALMI software \cite{shi2022}, proving to have a high rate of additional event detection while also having a low false-positive (fake events) detection.

\section{Data \& Study Area}
We used data collected as part of the COSEISMIQ project \cite{grigoli2022} aimed to help geoscientists to develop innovative tools for monitoring and controlling induced seismicity (Fig.~\ref{mapgraph}). The dataset includes a variety of seismic recordings that provide insights into the subsurface processes associated with geothermal energy production. The monitoring efforts were designed to capture the seismic response to geothermal activities, which is essential for understanding the relationship between human activities and seismic events. The collected data can help identify patterns of induced seismicity, which is crucial for mitigating risks associated with geothermal energy projects.
The dataset was collected after the deployment of a temporary seismic network in the Hengill geothermal region of Iceland. The area hosts the two largest geothermal power plants in the country are located:  the Nesjavellir ($\sim$\SI{120}{\mega\watt}) and Hellisheidi ($\sim$\SI{300}{\mega\watt}) power plants (the latter being the third largest EGS in the world) \cite{gunnlaugsson2016}. They both experience natural and injection-induced seismicity, with several thousand earthquakes per year and magnitudes exceeding $M_{w}$ $4$.
Continuous three-component waveforms were recorded from 1 December 2018 to 31 January 2021. Broadband and short-period nodes operate at 200 Hz; IMO stations run at 100 Hz. Raw miniSEED files and full StationXML metadata are openly distributed through FDSN webservices at the European Integrated Data Archive (EIDA) under network codes 2C, 4Q, OR, VI. All temporary 2C data are hosted permanently at the ETH-EIDA node; 4Q is archived at GFZ-EIDA; OR and VI are mirrored at ETH-EIDA.

The experiment produced both a 1D P and S seismic velocity models, followed by a 3D local earthquake tomography study \cite{obermann2022}. It also delivered 3 automatic seismic catalogs of different qualities and sizes, produced by implementing an ad-hoc SeisComp \cite{hanka2010, olivieri2012} system for automatic analysis both in \emph{playback} and \emph{real-time} mode
The three automated catalogues are classified by a score metric, named \emph{quality score} (S), that has been developed at the Swiss Seismological Service \cite[supplementary information]{grigoli2022, bagagli2022}. This score condenses azimuthal gap, phase count, RMS, source-station distance, and pick residuals into a single (negative) metric. To belong to any of these catalogs, an event must have at least $10$ picked phases associated. The high-quality (HQ), medium-quality (MQ), and low-quality (LQ) catalogs are separated by a subclassification of the score: $S \geq -1$, $-5 < S < -1$, and $S \leq -5$, respectively.
The progressively stricter score thresholds reduce the catalogue size from $\sim$\num{12000} (LQ) to $\sim$\num{9900} (MQ) and $\sim$\num{8500} events (HQ) while markedly improving location reliability, making the HQ set suited for precise seismotectonic interpretation, the MQ set useful for broader statistical studies, and the LQ set principally for completeness tests or machine-learning benchmarking where noisy labels are acceptable.

For our study, we collected the waveform data obtained only from the 2C and OR seismic networks. The stations of these networks include STS-2, LE-3D5s, and CMG-6T broadband sensors, sampled at 200 Hz, for a total of \num{36} velocimeters, \num{3} components, that will represent our graph's nodes (Fig.\ref{mapgraph}). We also utilize the manually compiled seismic catalog from the ISOR agency and the LQ, HQ catalogs from the COSEISMIQ experiment for the same period to compile the training dataset and compare our results, respectively.
For a full list and description of the temporary seismic networks and the COSEISMIQ experiment, we refer the reader to the manuscript of \citeA{grigoli2022}.

\section{Methods}

\subsection{Heimdall Architecture}
\begin{figure}
  \includegraphics[width=\columnwidth]{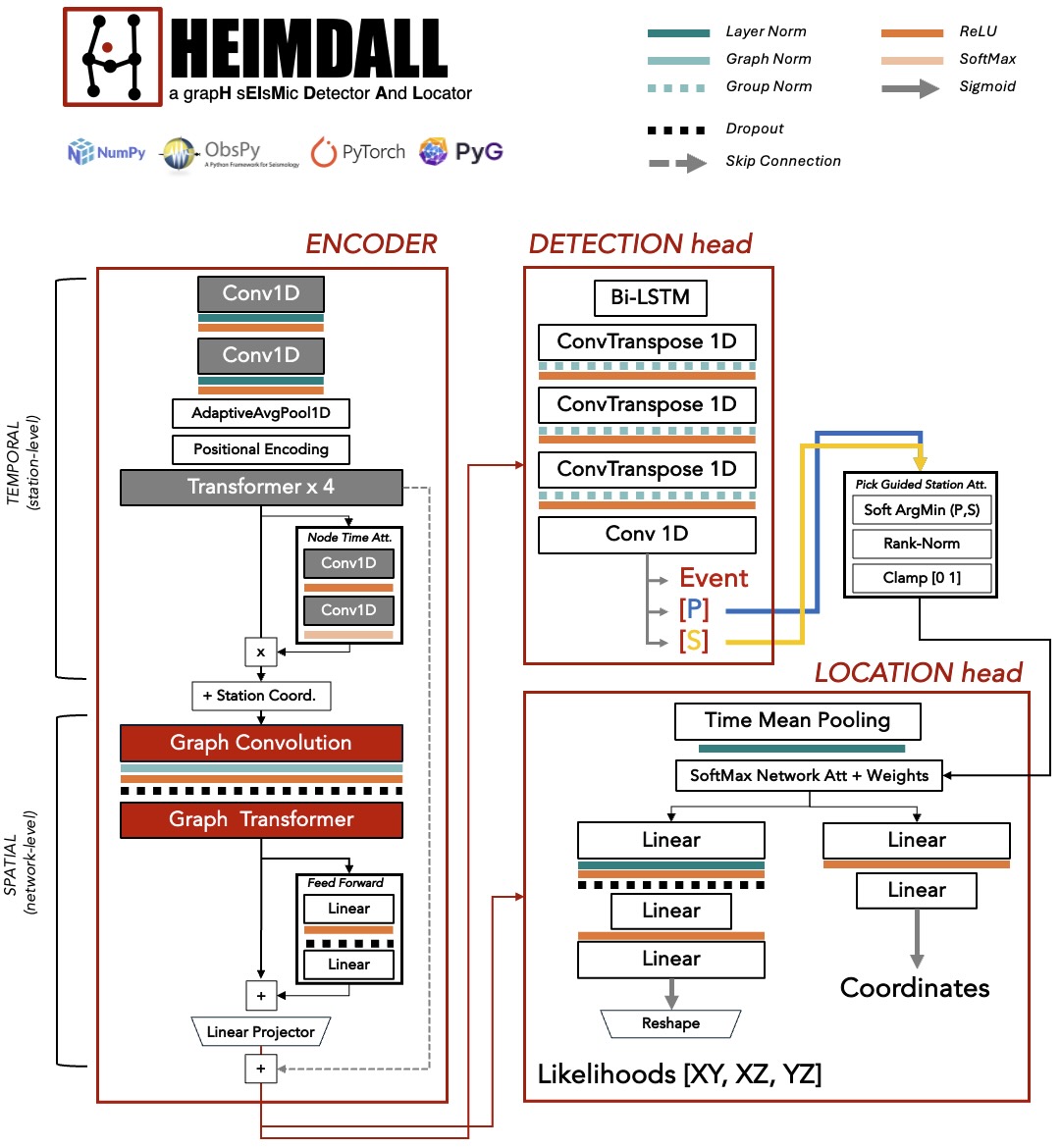}
  \caption{Heimdall architecture’s components breakdown. The encoder-decoder architecture ensures that each component of the software is specialized for a specific task: the encoder is responsible for describing the overall seismic network and inter-stations dependences, the detection head for the classification of event, P- and S- phases labels, and the locator head to produce the 2D projections of the event likelihood in space. For details, please refer to the methods section.}
  \label{heimarch}
\end{figure}

Heimdall is a deep learning model designed for microseismic event detection, phase picking, phase association, and event location within a single framework. It integrates multiple architectural components, including convolutional neural networks (CNNs), transformers, graph neural networks (GNNs), and attention mechanisms, to process continuous seismic waveforms from an array of stations (Fig.~\ref{heimarch}).

The model is a rather typical encoder-decoder structure that transforms raw multi-component seismic waveforms (with minimal pre-processing) recorded at a heterogeneous station network into (i) phase-consistent event probability traces, (ii) three dense spatio-probabilistic location maps, and (iii) a direct regression of hypocentral coordinates. Its design follows a strictly modular, yet fully differentiable, pipeline in which each stage is tailored to one of the three key axes of the problem (temporal, spatial, and graph-topological) before fusing them in a shared latent space.

While the encoder is responsible for extracting valuable information at a spatiotemporal level, there are \num{2} specialized heads that receive the shared encoder's embeddings and focus on their specific tasks: the \emph{Detector} head responsible for the 3 -labelling event, P-phase, and S-phase-; and the \emph{Locator} head responsible for producing 2D plane projections of the point-source likelihood placement.

\subsubsection{The Encoder body}
The backbone of the algorithm, responsible for the heaviest part of feature extraction and modality disentanglement, is the shared Encoder (from now on ENC) architecture (Fig.~\ref{heimarch}). This component operates in two sequential stages, temporal and spatial, that are carefully designed to isolate and then integrate the salient features of continuous waveform data. The encoding process is applied uniformly across all stations in the array, ensuring that learned representations are station-invariant while still allowing for local variability.

The temporal block operates on the raw multichannel waveform data at each station independently. Input traces are first passed through two successive 1D convolutional layers, each followed by ReLU activations and layer normalization. Layer-normalisation, preferred over batch-normalization, avoids batch-size artefacts, and ReLU activations \cite{krizhevsky2017} are applied after every convolution. These convolutions serve to extract channel-wise, short- and mid-term temporal patterns, capturing the signal morphology critical for discriminating between noise, P-wave, and S-wave arrivals. To mitigate overfitting and promote feature invariance to local shifts, a dropout layer \cite{srivastava2014} is applied subsequently. This initial convolutional stack reduces the temporal dimension and increases the representational capacity through feature expansion.

Following the convolutions, a temporal compression step is performed using an adaptive average pooling mechanism, which standardizes the time axis to a fixed length regardless of input duration. This operation preserves the dominant temporal features while enabling batch-level consistency and computational efficiency downstream. The pooled features are then passed to a series of 4 temporal Transformer encoders composed of multi-head self-attention layers and feed-forward networks. This module captures long-range dependencies within each station’s waveforms, allowing the model to infer global temporal context, such as the relative timing of wave arrivals, without requiring explicit windowing, thanks also to the spatial embeddings that treat each feature of the compressed time-vector as an individual token.
Before reaching the spatial block, the output from the transformer block undergoes a small self-attention pooling process. This step aggregates the data, enabling the network to identify the most critical time steps for the subsequent stages. As a result, this acts as an intelligent pooling layer, enhancing the network's performance.

The spatial block operates on the temporally encoded features of the entire seismic network. Station-level embeddings are assembled into a (densely) weighted graph whose nodes correspond to stations and whose edges are weighted by geodesic distance (see next section). A Graph Convolutional Network layer (GCNConv; \citeA{kipf2017}) first propagates information among neighbouring stations, contextualising local detections with regional patterns. The resulting node features are then refined by a Graph Transformer layer (TransformerConv; \citeA{shi2021}), which applies cross-station self-attention to capture both topological and long-range spatial correlations.

To remain as close as possible to the standard Transformer design \cite{vaswani2017}, we append a position-wise feed-forward network (FFN) to the output of the graph-Transformer encoder. We intentionally omit an additional residual connection and layer normalization at this stage, both of which are already present earlier (e.g., graph normalization after the GCN) to prevent overregularization. A final bias-free fully connected layer reshapes the feature tensor so it can be added to the temporal embeddings produced by the temporal Transformer, giving the encoder a holistic view of the network’s spatio-temporal structure while preserving the original temporal information.

This hierarchical encoder, comprising local 1-D convolutions, temporal self-attention, spatial message passing with Graph Convolutional Networks (GCNs), and topological self-attention with a Graph Transformer, produces a compact and information-rich latent space. That shared embedding feeds multiple task-specific decoder heads for detection, picking, and location, enabling each task to benefit from the common upstream representation while being optimised independently.

\subsubsection{The Detector head}
The detector head (hereafter referred to as DET) is the first of the three decoder modules. It is responsible for identifying the temporal occurrence of seismic events within continuous waveform data. Operating on the shared latent representation produced by the encoder, this head outputs a probability trace over time for each station, for all 3 classification channels: event duration, P-phase, and S-phase. \cite{ross2014}

Specifically, the latent features in the output from ENC are first reshaped to reintroduce the batch and station structure, along with the compressed temporal length, and are the input into a two-layer bidirectional LSTM (Bi-LSTM, \citeA{hochreiter1997}). Operating independently on every station’s sequence, the Bi-LSTM captures long-range dependencies in both forward and backward directions: the forward pass aggregates information from the past, while the backward pass looks ahead. Because the LSTM re-uses its hidden state at each time step, it can keep track of slowly building energy and bridge gaps caused by small glitches or missing samples, phenomena that typically challenge purely convolutional decoders DL pickers. Operating independently for every station ensures that the spatial context, already embedded upstream, is not overwritten.

The attention-refined embeddings are subsequently processed through a small stack of temporal convolutional layers, which serve as a local decoder for reconstructing per-station events and for classifying P- and S-phase time series. Following this, a sigmoid activation function \cite{narayan1997} is applied, producing continuous probability scores in the range of $[0,1]$ and resulting in a final tensor matching the original input time-series length.

The first channel, (\text{event}), represents the probability of an event window. P- and S-phase probabilities are \emph{soft-gated} by \text{event}:
\[
\text{P}(t)={P}(t)\times\text{event}(t), \quad
\text{S}(t)={S}(t)\times\text{event}(t),
\]
enforcing the physical requirement that every pick belongs to an event, and those cannot be found outside an event declaration.
This output can be interpreted as a binary classification for each node over time and is trained using Binary Cross-Entropy Loss \cite{mao2023} against annotated event, P- and S-windows.

To enhance the connection between phase picking, event detection, and event location, we employ a small pick-guided station attention block to process the P- and S-classification channels. This lightweight, fully differentiable gating module uses the P- and S-phase probability traces to inform the locator about which stations should receive more attention.
Initially, each trace is transformed into a continuous pick time using a soft-argmin operation. In this process, the time axis is treated as discrete support, the probabilities act as weights, and their normalized first moment produces a differentiable estimate of the pick time. Since this expectation is calculated independently for each station, gradients can still propagate back through the entire detection system.
Next, the pick times are linearly rescaled within each graph so that the earliest arrival is assigned the highest score, while the latest receives the lowest score. This approach reflects the physical intuition that stations closer to the hypocenter provide more accurate location information. The weights derived from the P and S phase estimates are combined and limited to a range of [0, 1], resulting in a single saliency coefficient for each node.

By framing detection as a node-wise temporal segmentation task, the DET harnesses both local signal characteristics and contextual information from neighboring stations (via the ENC) to achieve robust performance across varying noise and signal conditions.

\subsubsection{The Locator head}
The locator head (hereafter referred to as LOC) approach begins by utilizing the shared encoder output, specifically the temporally and spatially integrated latent tensor. A learnable pooling operation is applied across the temporal axis to compress the temporal dimension into a fixed-size summary for each station, resulting in a station-wise embedding that captures the event dynamics.

Next, a lightweight attention gate determines the influence each station has on the final location estimates. This is achieved by multiplying the weights from the pick-guided attention block to create a stronger connection with the remaining time-domain features.

These station-level descriptors are then aggregated using a permutation-invariant graph pooling layer. This allows the following decoder to work with a single, fixed-length vector, regardless of network density, which serves as the starting point for spatial decoding.

At this stage, the pooled embeddings are sent through two distinct gradient flow paths, both of which utilize a multi-layer perceptron (MLP). One path expands the embeddings into three planar projection images, while the other is responsible for providing the most probable 3D coordinates of the location. The former includes a low-rank layer to prevent parameter explosion in the final fully connected layer, which is then reshaped to create the three images. The latter, instead, is responsible for guiding the network towards physically coherent estimates, which are then regularized with an $L_{1}$ penalty against catalogue hypocenters. This design aims to guide the network towards physically coherent estimates, mitigating also bias introduced by uneven station geometry or phase-picking errors.

During training, the primary objective is to minimize the per-pixel binary cross-entropy (BCE) between the predicted and target likelihood images. Concurrently, an auxiliary regression loss provides a low-variance gradient that accelerates convergence and stabilizes the early stages of learning. At inference time, the coordinate head is not used directly to estimate the final event location; instead, the system identifies the peak of the likelihood image as the estimated source position, while retaining the entire image to represent epistemic uncertainty (see further in the text).

\subsection{Experiment setup}
To work, the Heimdall algorithm needs to have a graph (for analysis) and a grid (for location) to operate with. A graph,
$g = (\mathcal{V}_g, \mathcal{E}_g, \chi_g, A_g)$,
is defined by a set of vertices $\mathcal{V}_g$ (also referred to as nodes) and by a set of edges (or arcs) $\mathcal{E}_g$ connecting pairs of nodes \cite{bondy1976}. The latter specifies which (and how) nodes are interconnected in the graph, and this structural information is also known as \emph{adjacency matrix}. For a complete definition of a graph and its applications in the machine learning world, we refer the user to \citeA{bondy1976, bacciu2020} and the references therein.

For our case study, similarly to previous works (i.e, \citeA{bloemheuvel2023, mcbrearty2023}), we assimilate the network's seismic station to the nodes of our structure, and the edges as the spatial connections among those. Both nodes and edges contain features and attributes used in the processing. At the node level, we utilize the three minimally processed raw components (ZNE channels) of seismic records, which involve removing the mean, linear trends, and scaling to the range of $[-1,1]$. As edge attributes, we calculate the geodetic distances of the connection's lengths to better leverage the potential of the graphs in dealing with non-Euclidean problems. Euclidean data, such as images, audio signals, or videos, is defined on regular grids where standard notions of distance, angle, and neighborhood are consistent throughout the space. Convolutional Neural Networks (CNNs) are highly effective in these domains due to their ability to exploit local stationarity and translation invariance through shared convolutional filters. On the other hand, Graph Neural Networks (GNNs) extend the capabilities of deep learning to these non-Euclidean domains by learning representations that respect the underlying graph topology. This makes GNNs suitable for tasks where relationships between entities are more naturally captured through edges rather than spatial proximity.

The construction of graph edges remains a critical and unresolved challenge, and several algorithmic methods can be found in the literature for deciding how to build the connections. For example, the \textit{k}-nearest neighbors (KNN), where each node is connected to its closest \textit{k} neighbors in feature or spatial space \cite{cover1967}. Other approaches include density-based clustering algorithms such as DBSCAN \cite{ester1996}, which aim to capture local density variations by connecting nodes in high-density regions, potentially reducing spurious edges in sparse areas. More recent advances involve learnable edge construction mechanisms, where the connectivity is dynamically inferred during training using attention mechanisms or parameterized similarity functions. However, the question of how to construct edges most effectively remains a matter of ongoing debate within the community.

After several tests, we opted to construct the adjacency matrix as follows: (i) undirected, ensuring the matrix is symmetric; (ii) based on the K-nearest neighbors (KNN) approach, allowing each station to connect to at least 7 neighboring stations; (iii) using a static graph; and (iiii) including self-loops to allow each station to contribute to its own representation during message passing (Fig.\ref{mapgraph}). Self-loops are essential in Graph Neural Networks to preserve a node's intrinsic features and ensure that its state can influence its future updates. While the physical problem of seismic event localization traditionally requires a minimum of four observations (three to determine the spatial location in 3D and one for the origin time), grouping additional observations can significantly improve the accuracy and stability of the solution, especially in the early stages of model inference.

For the event location, Heimdall employs a fully parameterizable mesh grid that operates independently of the graph. This approach offers flexibility in sampling across all three dimensions and allows for construction in a specific area that is significantly smaller than the overall graph's aerial distribution. In our case study, due to the prevalent microseismicity, we chose a spacing of \SI{100}{\metre} throughout the network, while adjusting its extent to accommodate the investigation area, which measures \SI{30}{\km} by \SI{33}{\km} by \SI{15}{\km}.

\subsection{Training}
\begin{figure}
  \includegraphics[width=\columnwidth]{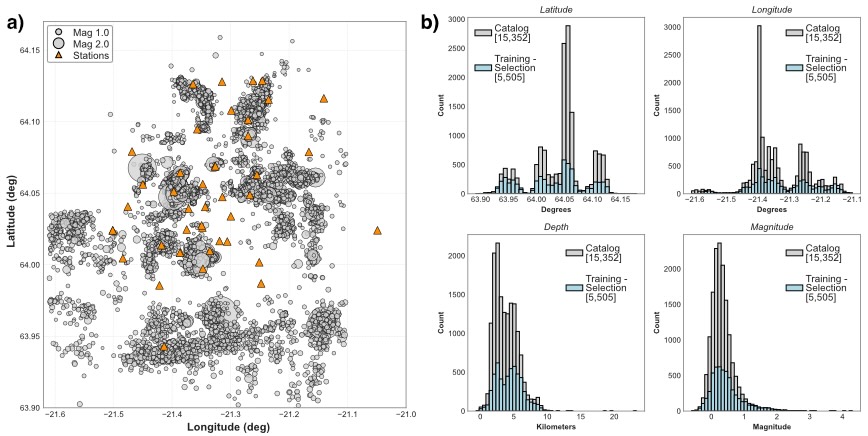}
  \caption{Training dataset selection. (a) 4 years of seismic events recorded by the COSEISMIQ network. (b) The histograms show the distribution of both the entire COSEISMIQ dataset (in grey) and the representative training dataset selected for this study (in light blue). To limit class imbalances, we chose to include all events with $M_w$ greater than 2.}
  \label{traindata}
\end{figure}

To construct the training dataset, we selected a representative subset of events from the complete list of manually repicked events during the COSEISMIQ experiment. From a total of \num{15352} events, we selected only \SI{33}{\percent} of them, including all the events with reported $M_{w}\geq2$ to limit the class imbalance of the dataset. This representative selection led to a total of \num{5505} events (Fig.\ref{traindata})

Because of the multi-tasking approach of Heimdall software, we needed to focus on deciding: (i) the input waveforms preprocessing, (ii) the DET labels, and (iii) the LOC labels.

For the input pipeline (i), we choose to: collect the necessary MSEED files, merge the stations' channels by filling the gaps with zeroes, and unify the sampling rate by downsampling all stations to \SI{100}{\hertz}. As the Heimdall software is meant to work on continuous data, using a pre-determined sliding window, so to allow not only application in \emph{playback mode} (like the one we are presenting in this work), but also for near real-time monitoring applications.  Based on the average S-P arrival times distribution and the microseismicity characteristic of the area, we opted for a 5-second window sampled at \SI{100}{\hertz} and a sliding window of \SI{0.5}{\second}. Each waveform in the window is then normalized by its individual channel's standard deviation, but will ultimately be rescaled to a range of $[-1, 1]$ values before being entered into the Heimdall model.

For the DET targets (ii), we adopted a standard approach widely used in ML software for phase-picking purposes \cite{ross2018, zhu2019, mousavi2020}, utilizing Gaussian-shaped functions around the P- and S-picks, and a box-shaped function for event labeling. For picks, we used a Gaussian-shaped kernel whose size is a function of the actual weights (based on absolute timing error) provided in the ISOR catalog for the manual routine operators. For the event-related window's length we used the traveltimes for P ($P_{t}$) and S ($S_{t}$), thus calculating for each event-station pair the time from the $P_{t}$ to the $S_{t} + 0.5 \cdot (S_{t}-P_{t})$ (Fig.~F1a-b, supplementary).

For the LOC targets (iii), we wanted to provide a set of 2D projection solutions, similar to the posterior probability distributions. As we are using the full potential of a waveform-based, statistical approach in a way similar to other methods like backprojection, array beamforming and stacking on travel-time delays (based on the MUSIC algorithm, \citeA{schmidt1986}) or coherence STA/LTA waveform migration and grid searching, \citeA{grigoli2014} seismic locator algorithms. As shown in Fig.~F1a-b (supplementary), we purposefully introduce uncertainty using an isotropic Gaussian kernel, again centered on the actual location of the event within the window. This is to mimic that the more information (i.e., picks) in the station windows, the sharper and more accurate the solution. For the 2nd LOC head (coordinate derivation), we use the original location of the catalog, which represents the peak of our images. Indeed, every image is scaled toward the $[0,1]$ value range to accommodate the downstream loss-function calculation effectively.

To enhance the generalization capability of our approach, we adopted a data augmentation strategy. Data augmentation techniques are especially pronounced when the available training set is modest, a frequent scenario at early stages of temporary experiments, where well-labelled events are scarce and costly to curate. Additionally, augmentation has been shown to halve false-alarm rates and stabilise training on datasets that would otherwise be too small for deep architectures \cite{zhu2020}. By synthetically diversifying the input distribution, augmentation forces the network to learn class-defining invariances rather than memorising, for example, particular source-receiver geometries. In Heimdall, we therefore treat augmentation as an integral part of the training pipeline. Every mini-batch is transformed on the fly, ensuring that the model never sees the same waveform twice and effectively multiplying the size of the dataset.

The chosen transformations follow geophysical priors while preserving label consistency. We mainly draw additive colored noise whose power spectrum matches that of the local ambient field, and perturb amplitudes through both gain scaling and dynamic-range compression (i.e., logarithmic compression) to simulate instrument response variability. Each transformation is stochastic, parameterised by ranges derived from empirical statistics, and composed such that the underlying label (event time and location) remains valid. Together, these augmentations expose Heimdall to a spectrum of plausible recording conditions, resulting in markedly improved robustness at inference stages.

As a simultaneous multi-tasking software for phase picking, association, and event location, the Heimdall training stages must be able to accommodate this complexity. Therefore, we defined a tailored loss function that combines residuals from all heads and back-propagates them into the models, updating their weights using the unified information from the spatio-temporal integration and temporal phase-time arrival links.

\begin{equation}
\begin{aligned}
  \mathcal{L}_{DET} &= BCE(event,P,S), \\
  \mathcal{L}_{LOC} &= BCE(xy) + BCE(xz) + BCE(yz), \\
  \mathcal{L}_{COORD} &= BCE(coord)
 \end{aligned}
\label{eqloss_heads}
\end{equation}

\begin{equation}
  \begin{aligned}
  \mathcal{L}_{HEIM} &= \alpha\, \mathcal{L}_{DET}+\beta\, \mathcal{L}_{LOC}+\gamma\,\mathcal{L}_{COORD} \\
  \alpha &= 1.0,\; \beta = 1.0,\; \gamma = 1.0.
  \end{aligned}
\label{eqloss_heim}
\end{equation}

The composite loss in Eq.~\ref{eqloss_heim} couples Heimdall’s two primary tasks, event detection and hypocentral location, into a single optimisation objective. The detection term, $L_{DET}$, is the binary cross-entropy (BCE) applied jointly to the three probability traces produced by the detector head: the event window indicator (EE) and the phase-specific channels (PP and SS). This formulation penalizes false alarms and missed picks symmetrically, encouraging well-calibrated probabilities, both of which are critical for the downstream locator. The localization term, $L_{LOC}$, is likewise a BCE but is evaluated on the three orthogonal likelihood images (XY, XZ, YZ), each representing a 2D marginal of the 3D posterior over source coordinates. Treating the maps pixel-wise, the BCE provides dense supervision, allowing the network to learn not only the most likely epicentre but also a physically meaningful uncertainty field. The coordinate term $L_{COORDS}$ regularises an auxiliary coordinate-regression head that converts the likelihood images into continuous $(x,y,z)$ estimates via a differentiable soft-argmax (Fig.~\ref{heimarch}). This term supplies a low-variance gradient that accelerates convergence and stabilises the early phase of training, when the images are still diffuse. Eq.\ref{eqloss_heim} linearly combines these components with weights $\alpha$, $\beta$, and $\gamma$ weight multipliers. In our case, the three parts are defined with equal weights $alpha=\beta=\gamma=1$, reflecting an empirical finding that balanced gradients yield mutually beneficial representations. Thus, accurate picks help the locator focus on high-quality stations; sharp likelihood maps guide the detector toward time windows that maximise spatial consistency; and the coordinate head aligns both with the physical hypocentre. Together, the three losses promote a coherent spatio-temporal understanding, meaning that Heimdall learns mutually consistent representations in its shared encoder. This results in superior robustness and calibration compared with training the two tasks in isolation.

Moreover, our loss definition enables the user to specify which tasks should be prioritized (or, more accurately, assisted) for Heimdall to focus on. For example, cases may arise where one of the three tasks reaches a plateau too soon, or when a class imbalance (especially for locating clustered seismicity) leads to a subtle model collapse in generating the likelihood images, thereby altering the final location.

\begin{figure}
\begin{center}

  \includegraphics[width=0.7\columnwidth]{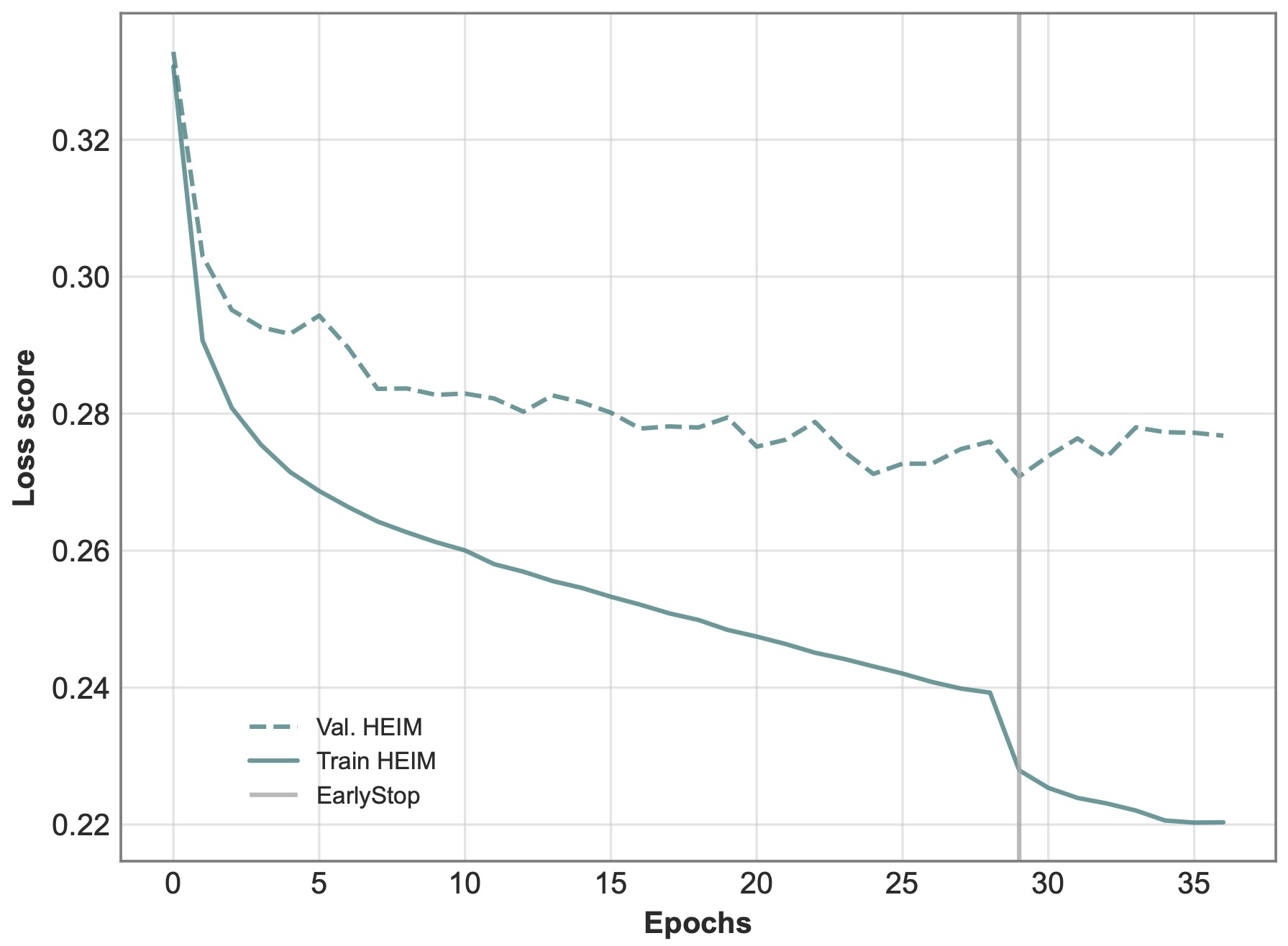}
  \caption{Training and validation loss curves. The early stopping criteria (Tab.~\ref{tab:hyperpar}) triggered at the 30th epoch, allowing a smoother convergence without the risk of overfitting the model. A reducing learning rate on plateau approach was used to increase the training epochs for a smoother transition in the later training stages.}
  \label{traincurves}
\end{center}
\end{figure}

The training results are shown in Fig.~\ref{traincurves}, indicating that our model converges toward a minimum. We used the hyperparameters reported in Tab. \ref{tab:hyperpar}. To fully leverage the model's learning capacity without overfitting, we employed an \emph{EarlyStop} approach combined with a scheduler to reduce the learning rate by a factor of 10 if the performance on the validation dataset falls below the given threshold. After numerous experiments, we found that this combination achieved the best balance between the DET and LOC head performances. The training lasted approximately \SI{20}{\hour} on a single GPU Tesla V100-PCIE-16GB, which was particularly fast compared to other seismicity DL models trained on multiple GPUs or with millions of training data elements.

\begin{table}[ht]
\centering
\caption{Heimdall's final hyperparameters used in the training stage.}
\label{tab:hyperpar}
\begin{tabular}{ll}
\hline
\textbf{Parameter} & \textbf{Value}\\
\hline
Random seed                   & 42\\
Noise windows          & \SI{10}{\percent}\\
Signal windows          & \SI{90}{\percent}\\
Train / Val / Test   & \SI{80}{\percent}/\SI{10}{\percent}/\SI{10}{\percent}\\
\hline
Otimizer                   & Adam\\
Learning rate              & $1\times10^{-4}$\\
Batch Size            & 8\\
Early stop:               & \\
\qquad Patience               & 7\\
\qquad Delta               & $1\times10^{-4}$\\
Scheduler  & ReduceLROnPlateau \\
\qquad Factor               & \num{0.1}\\
\qquad Patience               & \num{3}\\
\qquad Threshold               & $1\times10^{-3}$\\
\hline
$\alpha,\beta,\gamma$        & 1, 1, 1\\
\end{tabular}
\end{table}

\subsection{Inference stage}
Once trained, the Heimdall model can be applied directly to new data or in production mode to relocate seismicity. Conceptually, the software operates by sliding a 5-second time window at \SI{100}{\hertz} (501 samples) over the data. To declare the arrival of an event and proceed with source determination, a triggering method must be defined.

Therefore, we developed a rule-based classifier at the end of the Heimdall inference stage to analytically discriminate between signal-bearing and noise-dominated sets of three orthogonal plane projections (XY, XZ, YZ). This classifier performs a series of physically interpretable and statistically grounded checks to determine whether a given set of projections plausibly represents a localized seismic event:
\begin{enumerate}
    \item \emph{3D Spatial Consistency Check}. Given that the projections represent orthogonal slices of the same 3D seismic field, a genuine seismic event should result in consistent spatial localizations across all three projections. To evaluate this, the classifier computes the Euclidean spatial deviation among the three derived coordinates (XY, XZ, YZ). Specifically, it calculates the deltas between the X and Y coordinates from XY and XZ/YZ, respectively, and the Z coordinate from XZ and YZ, ultimately computing a 3D Euclidean spatial length. If this spatial delta, for each component, is below a user-defined threshold, the projections are deemed spatially consistent. We chose a threshold of \SI{4}{\km} for this check.

    \item \emph{Likelihood Sustainment Criterion}. To ensure that the PDF values across the projections are sufficiently strong, the classifier evaluates the maximum value in each plane and checks whether all three exceed a minimum threshold. This step ensures that the observed signal intensity is not trivially low in any projection, which might otherwise indicate noise or insufficiently localized energy. We chose a threshold of \num{0.05} for this check.

    \item \emph{Spatial Compactness via Variance}. True seismic events typically produce compact, high-energy regions in each projection. To quantify this, the classifier computes the variance of the PDF values in each projection. A low variance indicates that the energy is concentrated in a small region of the plane, aligning with expectations for a localized seismic source. The average of the three variances is compared to a user-defined threshold. If the average variance is below this threshold, the projections are considered spatially compact. We chose a threshold value of \num{3} for this check.
\end{enumerate}

The final binary classification decision, i.e., whether the projections represent an actual event or noise, is made by combining the three criteria above: spatial compactness, spatial consistency, and PDF sustainment. Only when all three conditions are simultaneously met is the set of projections classified as an event. Otherwise, the projections are labeled as non-event or noise. If five or more consecutive windows are deemed as event-signal, the triggers start to collect all the relative information until the next noise-signal window occurs. When that happens, Heimdall collects all the relative event windows and move on to finalizing the location.
For this final task, we implemented a robust multi-stage spatial localization strategy using all the plane-triplets collected. This method aggregates detection information across multiple time windows. It projects it on orthogonal planes (XY, XZ, YZ), from which a consolidated 3D hypocentral estimate is derived using statistical filtering, outlier rejection, and geometric triangulation.

\begin{enumerate}
    \item The first step involves extracting robust statistics from the sets of detection coordinates on each plane. The input event structure contains arrays of projected detection points over the XY, XZ, and YZ planes, each point represented by two spatial coordinates and an associated confidence or probability value. Each array is processed by computing a series of descriptive and robust statistics. A raw statistics: therefore, the mean, median, and median absolute deviation (MAD) calculations of the original coordinate set. An outlier filtering stage: thus, points deviating significantly (beyond 3.0 MADs) from the median are flagged as outliers. Clean statistics: where the mean, median, and MAD of the inlier subset are recalculated to provide the final, more reliable, spatial estimates. This step ensures that the downstream triangulation is based solely on high-quality, statistically representative localization points.

    \item \emph{Triangulation of 3D Hypocentral Location}. Cleaned detection coordinates from each plane are then passed to the triangulation function, which synthesizes a final 3D hypocentral estimate. This function estimates each spatial axis by averaging two independent sources: X is derived from the XY and XZ projections, Y is derived from the XY and YZ projections, and Z is derived from the XZ and YZ projections. 
        To quantify the coherence and confidence of the resulting location, the function aggregates the associated PDF values across all planes and computes the mean, median, standard deviation, and MAD. These metrics serve as volumetric analogs of detection certainty and solution compactness.

    \item \emph{Magnitude estimation}. Heimdall can also estimate the event magnitude by retrieving the maximum S amplitudes from the waveform cuts of each station. It can either calculate a relative magnitude ($M_{rel}$) if a dictionary with reference amplitudes from a known earthquake is provided, or estimate a local magnitude ($M_{L}$, \citeA{richter1935}) if a suitable attenuation function is given instead. We opted for the first solution by analyzing an $M_{w}=2$ recorded at the center of the seismic array, which was well received at all stations. We used the half peak-to-peak approach to minimize errors caused by spikes or sudden bursts of local energy (Text T1, supplementary information).
\end{enumerate}

This triangulation procedure reduces redundancy while leveraging multiple plane perspectives to stabilize the final estimate.
The derived 3D Cartesian coordinates are then converted to geographical coordinates (latitude, longitude) based on the original grid information, which maps internal grid coordinates to Earth-based locations. Additionally, the origin time of the event is coarsely estimated as the timestamp of the first detection window in the event sequence. While simplistic, this approximation is sufficient when fine-grained picking is deferred to later stages, or simply by using the picks in output from Heimdall's detection head within other software (e.g., NonLinLoc \citeA{lomax2000}, HypoDD \citeA{waldhauser2000}).

This modular approach not only yields a robust location estimate but also provides a diagnostic breakdown of detection quality across projections. The resulting information can be used to filter, rank, or further refine event detections in an automated seismic monitoring pipeline. Among all the possible detections that Heimdall provides, we chose to filter out all events that have a magnitude $< -1 M_{rel}$ and don't show at least 4 P picks (0.3 probability threshold).

\section{Results}

Once trained, we tested the Heimdall model on two different setups: a one-day experiment (2019-02-03) containing relatively low-frequency seismicity until a medium-size $M_{w}~2$ event occurred, generating a relatively standard seismic sequence; and a full month of data (December 2018) better to facilitate the comparison between the currently published ML catalogs: the SeisComp ones \cite{grigoli2022}, the one obtained by MALMI \cite{shi2022}, and the reference catalog in which the phase picks were manually reviewed and adjusted by ISOR (the ones from which we selected our training dataset). This is currently the most complete catalog in the analyzed time range and study area.

\subsection{The 2019-02-03 sequence}
\begin{figure}
  \centering
  \includegraphics[width=\columnwidth]{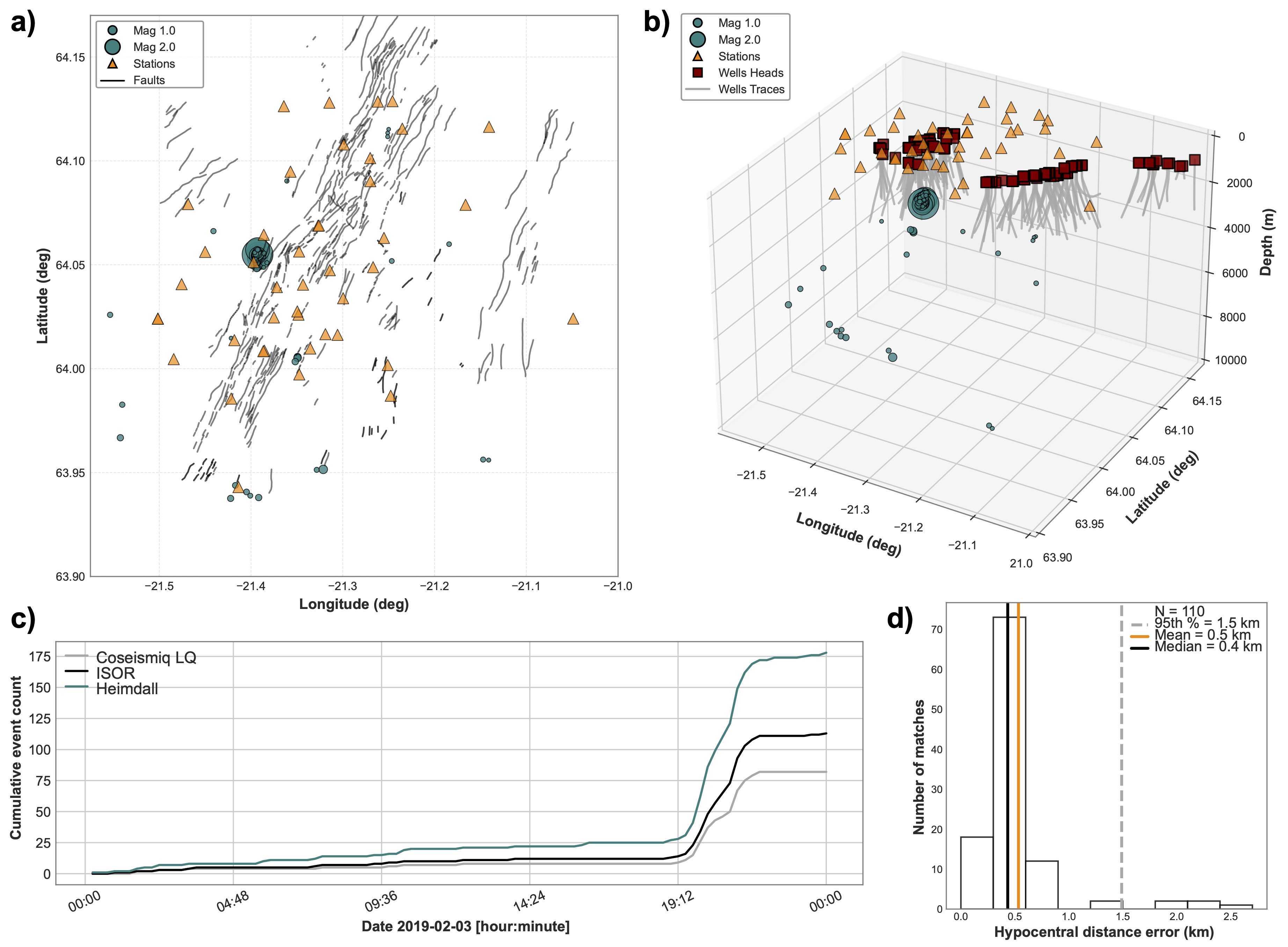}
  \caption{Results of the February 3rd seismic sequence: (a) 2D map showing the location of the seismic sequence; (b) 3D spatial distribution of the hypocenters; (c) Number of events detected and located compared to other catalogs (ISOR and COSEISMIQ); (d) Distribution of hypocenter location errors. The faults reported in panel (a) are extracted from \citeA{saemundsson1995}}
  \label{resultsFEB2019}
\end{figure}

\begin{figure}
  \centering
  \includegraphics[width=\columnwidth]{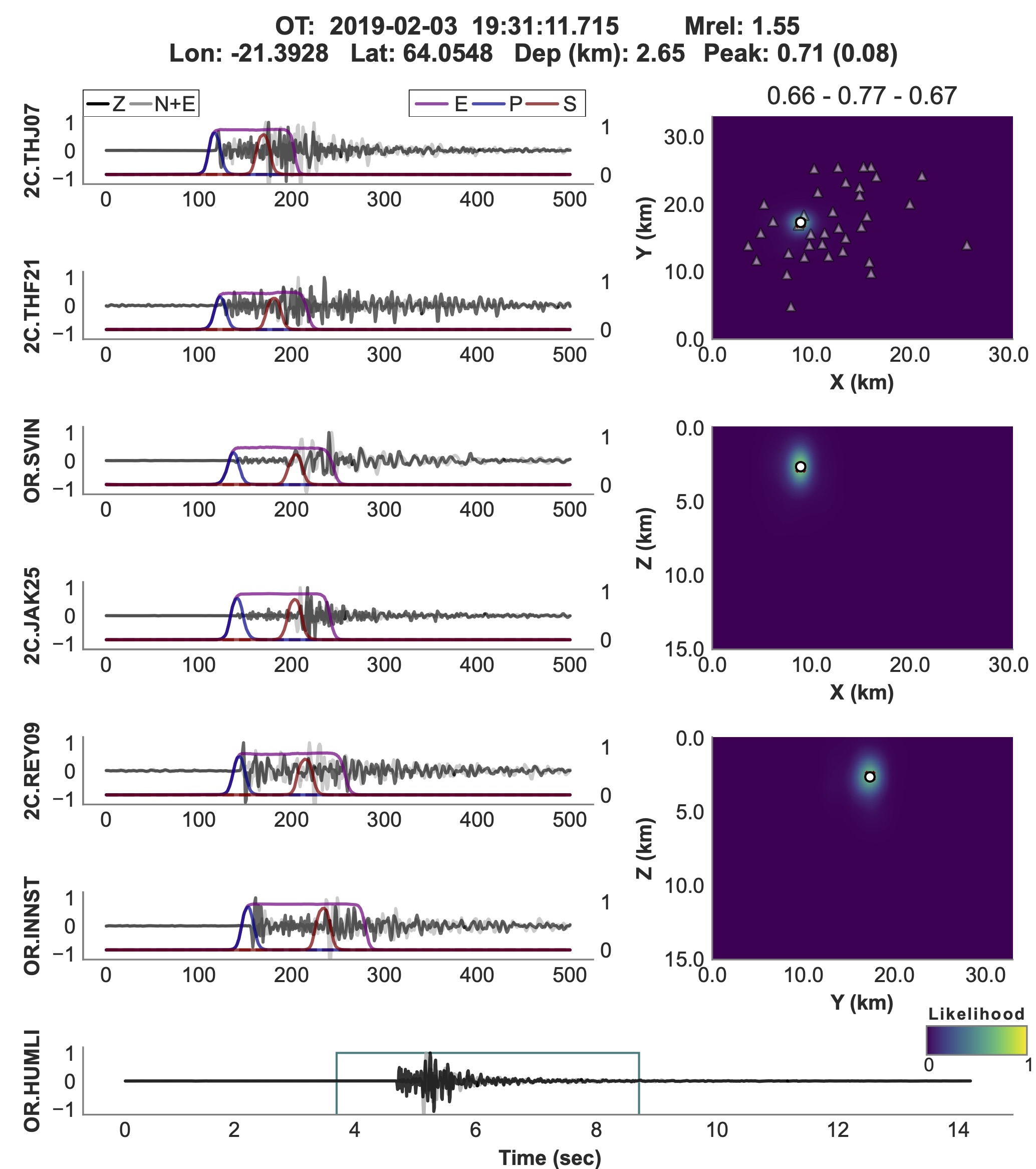}
  \caption{Heimdall workflow results example on a single window. The bottom time series shows the first triggered stations (closest epicentral distance from the final Heimdall's location), the light-blue square defines the window of interest; no label solutions are shown here to improve the figure's clarity. The time series plot in the left column shows the first 6 stations (from the 2nd closest station to the 7th closest one) and the relative detector's head outputs (event, P- and S- phases); real traces are normalized. The right panel shows the 3 images of the 2D source likelihood projections, respectively (from top to bottom) the XY, XZ, and YZ planes. Movie M1 in the electronic supplement shows the detection performance over continuous data. The white dot represents Heimdall's final location.
  }
  \label{resultsFEB2019_heim}
\end{figure}

The analysis of this sequence returned a total of \num{178} events from Heimdall (compared to the manual events retrieved in the catalog, for a total of \SI{\sim57.5}{\percent} increase in detections (Fig.~\ref{resultsFEB2019}). In Tab.~\ref{tab:statsresults} we provide some statistics obtained from manual investigation of the  2019-02-03 and 2018-12-30 seismic sequences (the latter better explained in the next subsection).
Heimdall's behavior, as anticipated in the previous section, involves rolling over a sliding window of 5 seconds sampled at \SI{100}{\hertz} (501 samples) over continuous data (Fig.~\ref{resultsFEB2019_heim}, movie M1 in the electronic supplement). This function is similar to a snapshot of the seismicity arriving at the recording stations, simultaneously picking up (if any) and locating events within the analysis grid. It can be used as a tool for reanalyzing a sequence or catalog to enhance resolution or increase the completeness of the catalog itself; however, it can also be viewed as a tool for near-real-time detection of microseismicity in the area of interest, for the network it was trained on.

To assess the correspondence between the manual and the detected automatic events, we implemented a pairwise matching algorithm based on temporal and spatial proximity. For each event in the manual reference catalog, we identified the closest event in the automatic catalog within a symmetric time window of \SI{15}{\second}, an epicentral distance threshold of \SI{5}{\km}. If multiple automatic events fell within the allowed time window, only the closest in time was retained. Moreover, to enforce a one-to-one association, each automatic event could be matched to at most one manual event. This strategy ensures high-resolution pairing while avoiding duplication of automatic detections. Notably, our matching criteria are more stringent than those used in prior works such as \citeA{grigoli2022} which adopted a \SI{30}{\second} and \SI{10}{\km} tolerance, and \citeA{shi2022} who applied a Chamfer Distance (CD, \citeA{park2021}) estimation only for the entire December 2018 catalog.

\begin{table}[ht]
\centering
\caption{Statistics relative to the manually revised sequence days results. We assume those average statistics as representative of the overall model performances.}
\label{tab:statsresults}
\begin{tabular}{c|cc|cccc}
\toprule
 Sequence Day & ISOR Events &  Missed & HEIM Events & Fake & Double report & Multi-event \\    
\midrule
 2019-02-03 & 113 & 3 (2.7\%) & 178 & 3 (1.6\%) & 2 (1.1\%) & 11 (6.2\%)\\    
 2018-12-30 & 142 & 3 (2.1\%) & 325 & 3 (0.9\%) & 4 (1.2\%) & 10 (3.1\%)\\    
\bottomrule
\end{tabular}
\end{table}

From the statistics, we note a significant decrease in the rates of missed events (False Negative, FN) and fake events (False Positive, FP), indicating the strong performance of Heimdall for this sequence. Although the rates of FN and FP are low, challenges arise in this area (i.e., microseismicity, low magnitude, short inter-event distances) where we occasionally have multiple events reported for the same detection (i.e., electronic supplement M1). However, these cases are relatively easy to identify when reviewing multiple picks at the same station. Indeed, in the methods section, we explained that there's a \emph{soft-gating} system that constrains the presence of P and S-phases only to an event classification, and these are usually reported as one event per-station, and per-detection (movie M2, electronic supplement). The reasoning behind this behavior lies in the final triggering and detection scheme. The precise separation of subsequent or short overlapping events in the automatic catalog (both with standard and ML approaches) is still difficult to fully address. For our method, the limitation is a 5-second sliding window that can still contain the coda of a previous event and the start of the subsequent one, which prevents the detection system from being classified as noise. These case scenarios avoid closing the detection window while maintaining momentum (electronic supplement M1).

In parallel with the overall cumulative distribution (Fig.~\ref{resultsFEB2019}c) showing a detection increase trend throughout the day, Fig.~\ref{resultsFEB2019}d presents the results from the hypocentral distance error matching comparison with the ISOR catalog. We observe that the mean and median values of the distribution are both approximately \SI{0.5}{\km}, thereby further demonstrating the consistency of our methodology.

\subsection{The December 2018 data}

We chose to analyze the entire December 2018 data as it is the only month in the area that is covered by multiple seismic catalogs produced with diverse algorithms (i.e., the SeisComp - COSEISMIQ, \citeA{grigoli2022}, the MALMI-DD one, \citeA{shi2022}. In this month, additionally, on the $30^{th}$ December, there was a $M_{w} 4.2$ in the southern-most part of the temporary seismic network (azimuthal GAP$>$\SI{180}{\degree}, one of the only 2 events with  $M_{w}\geq4$ for the entire COSEISMIQ project's period. This day is of particular interest given the complexity and high seismicity rate after such a mainshock.

As we did for the previous sequence, we also manually analyzed the Heimdall results for the $30^{th}$ December sequence (Table~\ref {tab:statsresults}, Fig.~F2 in supplementary). Even in this case, the pattern doesn't change: (i) we see a really low percentage of FN and FP, (ii) and still 10 multi-events reported in the duplicate detection (although this time, fewer in percentage to the overall higher detection rate). In total, Heimdall can detect \num{325} events compared to the \num{142} contained in the ISOR catalog.

\begin{figure}
  \centering
  \includegraphics[width=\columnwidth]{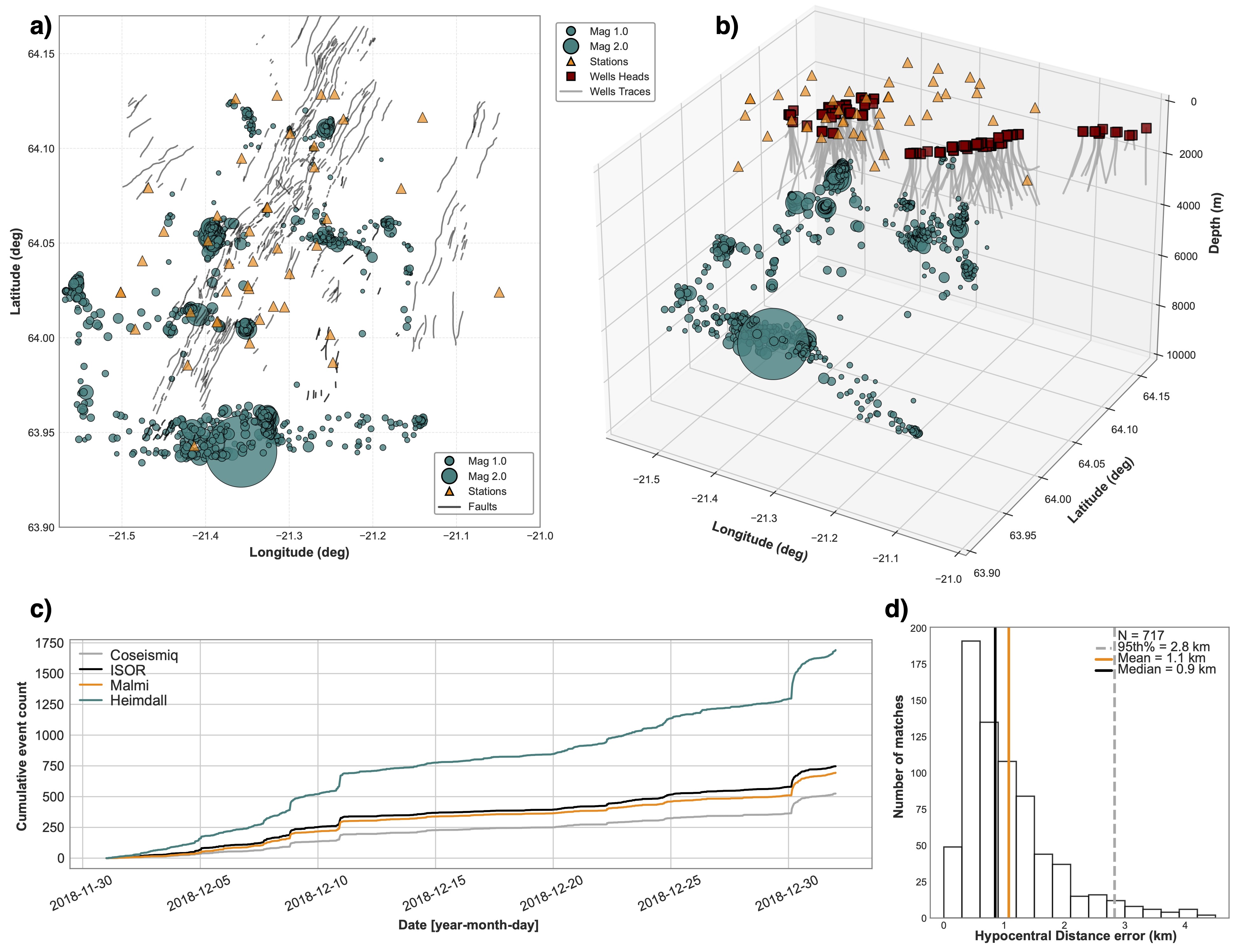}
  \caption{Results of the December 2018 analysis: (a) 2D map showing the location of the seismic sequence; (b) 3D spatial distribution of the hypocenters; (c) Number of events detected and located compared to other catalogs (ISOR and COSEISMIQ); (d) Distribution of hypocenter location errors. The faults reported in panel (a) are extracted from \citeA{saemundsson1995}}
  \label{resultsDEC2018full}
\end{figure}

\begin{figure}
  \includegraphics[width=\columnwidth]{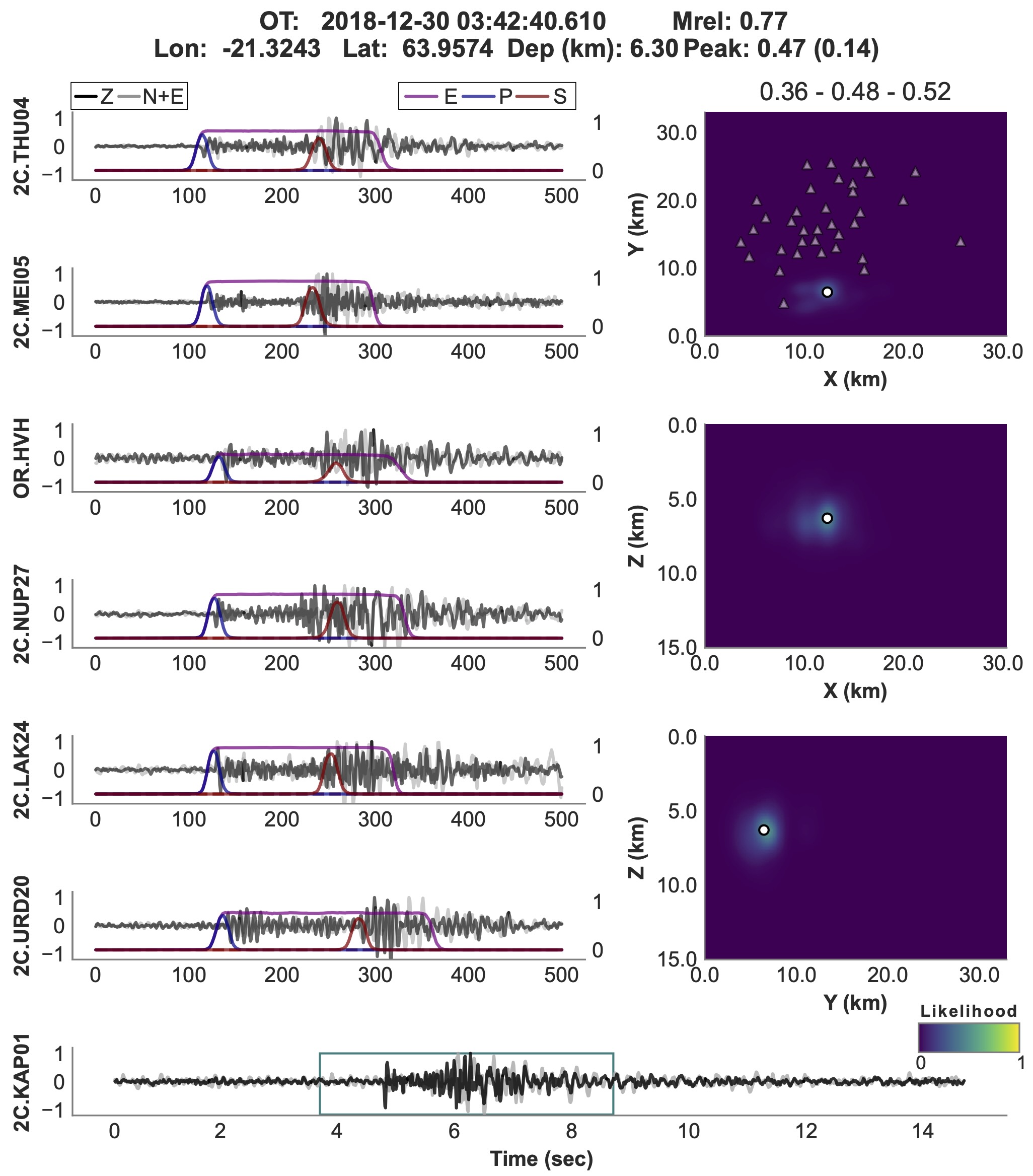}
  \caption{Heimdall workflow results example on a single window. The bottom time series shows the first triggered stations (closest epicentral distance from the final Heimdall's location), the light-blue square defines the window of interest; no label solutions are shown here to improve the figure's clarity. The time series plot in the left column shows the first 6 stations (from the 2nd closest station to the 7th closest one) and the relative detector's head outputs (event, P- and S- phases); real traces are normalized. The right panel shows the 3 images of the 2D source likelihood projections, respectively (from top to bottom) the XY, XZ, and YZ planes. Movie M3 in the electronic supplement shows the detection performance over continuous data. The white dot represents Heimdall's final location.}
  \label{resultsDEC2018full_heim}
\end{figure}

Regarding the entire month's results, Heimdall detects a total of \num{1691} seimic events, compared to the \num{747} events of the ISOR catalog (Fig.~\ref{resultsDEC2018full}). Overall, we observe a general trend from Heimdall to clusterize the seismicity effectively, while retaining the information about the South-Western seismicity of the area, especially where a low-angle dipping structure is much more prominent compared to the previous automatic catalog \cite{shi2022, grigoli2022}. The production example in Fig.~\ref{resultsDEC2018full_heim} (and in movie M3, electronic supplement) shows a weaker, low-magnitude seismic event located at the edges of our seismic network (higher azimuthal gap), where Heimdall still accurately locates the events even under such conditions in terms of phase-picking, and association.

\begin{figure}
  \centering
  \includegraphics[width=\columnwidth]{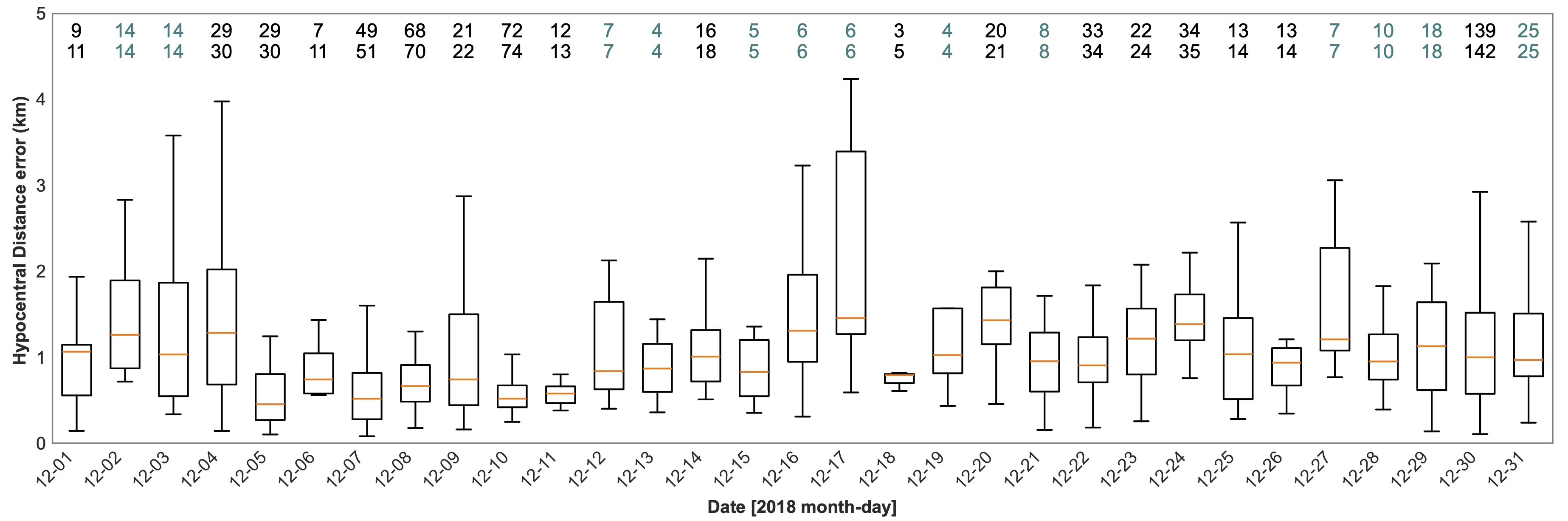}
  \caption{Daily box-plot of the hypocentral error distribution between Heimdall final catalog and the reference ISOR (manual) one for the entire December 2018 analysis.}
  \label{resultsDEC2018full_day}
\end{figure}

The entire month's matching result is shown in Fig.~\ref{resultsDEC2018full}d: we see a slight worsening of the average hypocentral difference (around \SI{1}{\km}). In Fig.~\ref{resultsDEC2018full_day} we split the full comparison into single-days box-plot statistics, to help the visualization of performances throughout the month. Overall, the mean value of the daily-distributions doesn't diverge too much from the overall mean and median values (as should be expected). In total, Heimdall is missing only the \SI{4}{\percent} of the reference catalog for the December 2018, while gaining a \SI{126}{\percent} event increase, definitely an improvement from the previous automatic catalog applied in the region.

A more quantitative, widely used parameter for evaluating the similarity between earthquake catalogs, and their event location distribution, we calculated the Chamfer Distance (CD), which is a similarity evaluation metric for two point sets and defined as \citeA{park2021} (Eq.~\ref{chamfer}). In Eq.~\ref{chamfer}, $x_{i}$ and $y_{j}$ represent an event hypocenter location in catalog X and Y, respectively, and $N_X$ and $N_Y$ are the total number of events in catalog X and Y, respectively.

\begin{equation}
\mathrm{CD}(X, Y) =
\frac{1}{N_X} \sum_{x_i \in X} \min_{y_j \in Y} \|x_i - y_j\|_2^2 +
\frac{1}{N_Y} \sum_{y_j \in Y} \min_{x_i \in X} \|y_j - x_i\|_2^2
\label{chamfer}
\end{equation}

When comparing a low-resolution catalog with a high-resolution one, the distance should decrease as the resolution of the high-resolution catalog increases because 'resolving' in this context refers to distinguishing the same underlying structures. In this case, lower scores mean a more compact and complete catalog compared to the reference. The resulting scores are reported in Tab.~\ref{chamfer}. We observe that Heimdall performs significantly better than the other automatic catalogs. This is most likely because we used real data extracted from the ISOR catalog itself, which helps in learning the original velocity model. In contrast, SeisComp catalogs and MALMI used the Minimum 1D reported in \citeA{grigoli2022}. Still, its high scores (calculated during inference) demonstrate Heimdall's potential for analyzing thick, intense seismic sequences without losing clustering capabilities.

\begin{table}[ht]
    \centering
    \caption{Chamfer distance scores between ISOR and other automatic catalogues for the same period. We can appreciate the lowest score reached by Heimdall (best performance) over the previously published catalog in the area: MALMI-DD \cite{shi2022}, and COSEISMIQ ones \cite{grigoli2022}.}
    \begin{tabular}{lcccc}
        \toprule
         & HEIM & SCP-HQ-DD & MALMI-DD & SCP-LQ \\
        \textbf{Chamfer distance (ISOR) Epi.} & 0.50 & 1.07 & 1.22 & 2.89 \\
        \textbf{Chamfer distance (ISOR) Hypo.} & 1.08 & 2.23 & 2.76 & 24.11 \\
        \bottomrule
    \end{tabular}
\end{table}

Eventually, we pursue a simplistic and relatively straightforward Frequency Magnitude Distribution (FMD) to fit the Gutenberg-Richter Law (GR, \citeA{gutenberg1944}), estimating the Magnitude of completeness and b-values for the December 2018 data (Fig. F3). We used the robust Lilliefors test (LF, \citeA{lilliefors1967}) to estimate the magnitude of completeness, resulting in a value of 0.3. The associated a and b-values are calculated with a bias-free maximum-likelihood regression \cite{tinti1987, marzocchi2003}, resulting in 3.29+/-0.01 and 0.96 +/-0.03 respectively. Although not strictly an absolute magnitude scale (i.e., $M_w$ or $M_L$), the relative magnitude can still provide a rough, first-order indication of the completeness of our catalog and allow for comparison of magnitudes within our catalog.

\section*{Discussion and Conclusion}

This work demonstrates that Heimdall delivers state-of-the-art performance while removing many of the operational bottlenecks that have long characterised conventional pick, associate, and locate pipelines.  By collapsing picking, association, and location into a single trainable architecture, Heimdall eliminates the need to tune disparate hyperparameters across multiple software packages and sidesteps error propagation between sequential stages. A complete training cycle on a single modern GPU finishes in \SI{\sim24}{\hour}.  Even a consumer-grade laptop equipped with an integrated GPU can scan one month of waveform data in roughly $\sim1.5$ days, underscoring its suitability for rapid-response deployments and field campaigns with limited computational resources and lightweight portability.

Heimdall’s flexibility stems from two design choices. First, its encoder produces a shared latent representation that is consumed by task-specific heads. Users interested exclusively in phase picking or event detection may train and deploy the corresponding head in isolation, thereby reducing both the memory footprint and inference time.  Second, the graph and grid domains are decoupled: station connectivity (edges) and hypocentral discretisation may be defined independently to target, for example, a specific fault system with dense graph edges while retaining a coarse grid elsewhere.  This decoupling naturally supports scalable experiments from single reservoirs to regional networks and paves the way for adaptive grid refinement.

Notwithstanding these advantages, Heimdall shares the intrinsic limitations of supervised learning.  The size, diversity, and heterogeneity of the labelled dataset constrain its generalisation capability.  Although class imbalance is mitigated by our representative sampling of the available catalogue, the model cannot be transplanted to a tectonically distinct region without some degree of retraining.  Clustered seismicity further exacerbates this issue because the network tends to over-fit to the dominant source clusters present in the training set. A good-quality, albeit modest in size, pre-existing catalogue, therefore, remains a prerequisite.

There are several ways to ease these constraints. We are experimenting with a self-supervised pre-training phase in which the encoder learns inter-station relationships and prevailing noise characteristics from a few days of raw recordings at a newly installed array.  Subsequent fine-tuning of the task heads with only a handful of hand-picked events can bootstrap a performant local model at the early stage after the station deployment. Coupling this workflow with scheduled continual-learning cycles will allow Heimdall to accommodate temporal changes in background seismicity or network geometry, thereby preserving accuracy without relying on a static, potentially outdated checkpoint. In parallel, a lightweight travel-time module exploiting the softmax attention weights of the P and S channels could also furnish rapid hypocentral estimates in poorly sampled sectors of the grid, much like classical table-lookup locators (i.e., NonLinLoc, HypoDD, LOKI), but with fully differentiable uncertainties. The limitation of assuming a single event at maximum in the same analysis window may affect the potential for reducing multi-triggered events. Modifying the time window's length during the creation of the training dataset may help close this gap; however, the classifier at the end also needs to implement new logic for handling multiple peaks in the images. Although not a problem in our case study (and possibly not in every local network), newer versions of the software would benefit from dynamic edge definitions to better constrain the solution based on actual data and reduce its impact on Heimdall's performance in the event of multiple stations shutting down simultaneously.
Two additional developments appear particularly promising.  First, embedding an eikonal-based solver within the locator head would enable explicit velocity model updates during training, closing the loop between waveform features and subsurface structure.  Second, the architecture is well-positioned to ingest high-density acquisitions such as distributed acoustic sensing (DAS).  Treating aggregated DAS channels as virtual nodes in the station graph would dramatically enhance azimuthal coverage (i.e., offshore or in wells deployment) while respecting the current computational envelope through dimensionality reduction.

To conclude, we presented a new spatio-temporal architecture tailored for continuous microseismic monitoring at a local scale, capable of re-processing offline pre-existing datasets and, once trained, enabling near-real-time monitoring applications for the specific network. The Heimdall algorithm is a comprehensive solution that performs phase detection, association, and event location in a single pass, providing uncertainties for all observations. In our site test in the Hengill region, we achieved good performance on two different setups (the entire December 2018 and the 2019-02-03 sequence), doubling event detections while achieving the best results on standard similarity tests, such as the Chamfer distance with the manually compiled ISOR catalog. By leveraging the potential for simultaneous picking and location, we also excelled at reducing false event detections and minimizing the loss of manually detected events. Heimdall leverages transformer-derived temporal features and graph-based spatial reasoning to deliver a unified, low-latency solution that complements conventional monitoring infrastructure for microseismic analysis at EGS sites or similar tectonic settings. Although region-specific retraining remains essential, the combination of rapid training, fast inference, and modular design positions Heimdall as a practical and extensible framework for next-generation seismic monitoring and reservoir surveillance.

%
%

\section*{Open Research Section}
The continuous seismic data used in this study are available at the Swiss Seismological Service (SED) eida nodes, with FDSN protocol network codes: OR and 2C. The earthquake catalogs used for comparison are stored here: \url{https://springernature.figshare.com/collections/Monitoring_microseismicity_in_the_Hengill_Geothermal_Field_Iceland/5725592} \cite{grigoli2022}, last accessed on June 2025. The Heimdall software and the related scripts for data preparation, training, and inference are also open-source and available on GitHub \url{https://github.com/mbagagli/heimdall}. The catalogs created in this work, along with the manual ISOR catalog, are stored in a permanent repository here: \url{https://doi.org/10.5281/zenodo.15804767}.

\section*{Declaration of Competing Interests:}  The authors acknowledge that there are no conflicts of interest recorded.

\acknowledgments
This work is part of the DERISK project supported by the European Union’s research and innovation program under the Marie Sklodowska Curie Grant (Agreement \num{101105516}). Many figures in the manuscript (and in the supplementary information) were created using the Matplotlib library \cite{hunter2007}. Fig.~\ref{mapgraph} was created using the Generic Mapping Tool (GMT; \citeA{wessel2019}) package. The seismic data processing and I/O have been done using the ObsPy library \cite{beyreuther2010, beyreuther2012}. The Heimdall model implementation was done using PyTorch \cite{paszke2019}, and PyTorch Geometric \cite{fey2019}.

\textbf{Author contribution statement:} Conceptualization: all authors; Funding Acquisition and Resources: FG; Methodology: MB, DB; Software: MB; Data Curation: MB, FG; Investigation: MB, FG; Visualization: MB; Writing original draft: MB; Writing review and editing: all authors.

\bibliography{BIBLIO}

\end{document}